\documentclass[11pt]{article}
\catcode`\@=11
\def\eqnarray{\let\@currentlabel=\theequation\refstepcounter{equation}
    \global\@eqnswtrue
    \global\@eqcnt\z@\tabskip\@centering\let\\=\@eqncr
    $$\halign to \displaywidth\bgroup\@eqnsel\hskip\@centering
      $\displaystyle\tabskip\z@{##}$&\global\@eqcnt\@ne 
       \hfil${{}##{}}$\hfil
      &\global\@eqcnt\tw@ $\displaystyle\tabskip\z@{##}$\hfil 
       \tabskip\@centering&\llap{##}\tabskip\z@\cr}
\def\lefteqn#1{\hbox to 4\arraycolsep{$\displaystyle #1$\hss}}
\long\def\@makefntext#1{\parindent 0cm\noindent
\hbox to 1em{\hss$^{\@thefnmark}$}#1}
\newcommand{\beq}{\begin{equation}}
\newcommand{\eeq}{\end{equation}}
\def\rref#1{(\ref{#1})}
\begin{document}
%
%
%
%
\def\citen#1{%
\edef\@tempa{\@ignspaftercomma,#1, \@end, }
\edef\@tempa{\expandafter\@ignendcommas\@tempa\@end}%
\if@filesw \immediate \write \@auxout {\string \citation {\@tempa}}\fi
\@tempcntb\m@ne \let\@h@ld\relax \let\@citea\@empty
\@for \@citeb:=\@tempa\do {\@cmpresscites}%
\@h@ld}
%
\def\@ignspaftercomma#1, {\ifx\@end#1\@empty\else
   #1,\expandafter\@ignspaftercomma\fi}
\def\@ignendcommas,#1,\@end{#1}
%
%
\def\@cmpresscites{%
 \expandafter\let \expandafter\@B@citeB \csname b@\@citeb \endcsname
 \ifx\@B@citeB\relax 
    \@h@ld\@citea\@tempcntb\m@ne{\bf ?}%
    \@warning {Citation `\@citeb ' on page \thepage \space undefined}%
 \else
    \@tempcnta\@tempcntb \advance\@tempcnta\@ne
    \setbox\z@\hbox\bgroup 
    \ifnum\z@<0\@B@citeB \relax
       \egroup \@tempcntb\@B@citeB \relax
       \else \egroup \@tempcntb\m@ne \fi
    \ifnum\@tempcnta=\@tempcntb 
       \ifx\@h@ld\relax 
          \edef \@h@ld{\@citea\@B@citeB}%
       \else 
          \edef\@h@ld{\hbox{--}\penalty\@highpenalty \@B@citeB}%
       \fi
    \else   
       \@h@ld \@citea \@B@citeB \let\@h@ld\relax
 \fi\fi%
 \let\@citea\@citepunct
}
%
\def\@citepunct{,\penalty\@highpenalty\hskip.13em plus.1em minus.1em}%
%
%
\def\@citex[#1]#2{\@cite{\citen{#2}}{#1}}%
%
%
\def\@cite#1#2{\leavevmode\unskip
  \ifnum\lastpenalty=\z@ \penalty\@highpenalty \fi 
  \ [{\multiply\@highpenalty 3 #1
      \if@tempswa,\penalty\@highpenalty\ #2\fi 
    }]\spacefactor\@m}
\let\nocitecount\relax  
%
\vspace{.5in}
\begin{flushright}
UCD-97-10\\
November 1997\\
\end{flushright}
\vspace{.5in}
\begin{center}
{\Large\bf
 Kinetic Energy\\[1ex] and the Equivalence Principle}\\
\vspace{.4in}
{S.~C{\sc arlip}\footnote{\it email: carlip@dirac.ucdavis.edu}\\
       {\small\it Department of Physics}\\
       {\small\it University of California}\\
       {\small\it Davis, CA 95616}\\{\small\it USA}}
\end{center}

\vspace{.5in}
\begin{center}
\begin{minipage}{4.2in}
\begin{center}
{\large\bf Abstract}
\end{center}
{\small
According to the general theory of relativity, kinetic energy 
contributes to gravitational mass.  Surprisingly, the observational 
evidence for this prediction does not seem to be discussed in 
the literature.  I reanalyze existing experimental data to test 
the equivalence principle for the kinetic energy of atomic electrons,
and show that fairly strong limits on possible violations can be
obtained.  I discuss the relationship of this result to the occasional 
claim that ``light falls with twice the acceleration of ordinary matter.''
}
\end{minipage}
\end{center}
\newpage

The principle of equivalence---the exact equality of inertial and
gravitational mass---is a cornerstone of general relativity, and 
experimental tests of the universality of free fall provide a large 
set of data that must be explained by any theory of gravitation.  But 
the implication that energy contributes to gravitational mass can be 
rather counterintuitive.  Students are often willing to accept the 
idea that potential energy has weight---after all, potential energy 
is a rather mysterious quantity to begin with---but many balk at the 
application to kinetic energy.  Can it really be true that a hot brick 
weighs more than a cold brick?

General relativity offers a definite answer to this question, but the
matter is ultimately one for experiment.  Surprisingly, while observational 
evidence for the equivalence principle has been discussed for a variety 
of potential energies, the literature appears to contain no analysis of 
kinetic energy.  The purpose of this paper is to rectify this omission, 
by reanalyzing existing experimental data to look for the ``weight'' of 
the kinetic energy of electrons in atoms.  I will then try to reconcile 
the results with the occasional (and not completely unreasonable) claim 
that ``objects traveling at the speed of light fall with twice the 
acceleration of ordinary matter.''

\section{The Equivalence Principle and Internal Energies \label{sec1}}

Modern tests of the principle of equivalence begin with variations of 
Galileo's apocryphal free fall experiment, in which the gravitational
accelerations of two objects with different compositions are compared.
The acceleration due to gravity is proportional to the ratio $m_G/m_I$
of gravitational to inertial mass, and violations of the principle 
of equivalence would manifest themselves as differences in acceleration, 
or ``nonuniversality of free fall.''  Some 
experiments${}^{\citen{Nie,Kuroda,Caru}}$ directly measure free fall; 
others${}^{\citen{Dicke,Moscow,EotWash}}$ use torsion pendulums to 
compare centrifugal acceleration to the acceleration due to gravity.  
Galileo's pendulum experiments${}^{\citen{Ciufolini}}$ may have already 
achieved accuracies of about $2\times 10^{-3}$, and since the pioneering 
work of E{\"o}tv{\"o}s,${}^{\citen{Eotvos}}$ modern experiments have 
pushed uncertainties down to about $10^{-12}$. 

These experiments directly test the equality of gravitational and
inertial mass for a variety of substances.  But because different 
materials have different compositions, the experiments also test the 
principle of equivalence for various forms of internal energy.  For
example, the inertial mass of an iron nucleus is about $1\%$ less than
the inertial masses of its constituent protons and neutrons, largely
because of the (negative) contribution of nuclear binding energy.  If 
this binding energy did not also affect gravitational mass, the ratio 
$m_G/m_I$ for iron would be greater than one, and iron would fall more 
rapidly than, for example, hydrogen. 

A convenient measure of potential violations of the equivalence principle 
is the E{\"o}tv{\"o}s ratio${}^{\citen{Will}}$
\beq
\eta(A,B) = {m_G(A)\over m_I(A)} - {m_G(B)\over m_I(B)} ,
\label{a1}
\eeq
where $m_G$ and $m_I$ are the gravitational and inertial masses of two
materials $A$ and $B$.  Typical experimental limits, which we shall use 
later, are 
\beq
\eta({\mathrm{Be}},{\mathrm{Cu}}) = (-1.9\pm2.5)\times10^{-12} 
\label{a2}
\eeq
for beryllium and copper${}^{\citen{EotWash}}$ and
\beq
\eta({\mathrm{Al}},{\mathrm{Pt}}) = (-0.3\pm0.9)\times10^{-12} 
\label{a2a}
\eeq
for aluminum and platinum.${}^{\citen{Moscow}}$
Now, the inertial mass of a material includes a number of 
contributions---rest mass, nuclear binding energy, electrostatic energy, 
the kinetic energy of constituents, etc.  If any of these internal 
energies $E_\alpha$ were to violate the principle of equivalence, one 
could write, to lowest order in the energies,
\beq
m_G = m_I + \sum_\alpha \eta_\alpha {E_\alpha\over c^2} ,
\label{a3}
\eeq
where the $\eta_\alpha$ are dimensionless parameters that measure the
strength of the violation.  From \rref{a1} and \rref{a3}, we see that
\beq
\eta(A,B) = \sum_\alpha \eta_\alpha \left({E_\alpha(A)\over m_I(A)c^2} 
  - {E_\alpha(B)\over m_I(B)c^2}\right) .
\label{a4}
\eeq
In the absence of fortuitous and fine-tuned cancellations among various 
types of energy, observational limits on $\eta(A,B)$ can thus be used 
to obtain limits on the parameters $\eta_\alpha$.  

Expression \rref{a3} is, of course, an oversimplification: in practice, we 
can rarely separate the internal energy of a material so cleanly into its 
components.  The binding energy of protons and neutrons in the nucleus, 
for example, is an important part of the energy of an atom, typically 
amounting to a bit less than $1\%$ of the mass.  But protons and neutrons 
are made up of quarks, and according to quantum chromodynamics, nuclear 
binding energy is merely a remnant of the interaction energy of quarks 
and gluons.  Should the binding energy in \rref{a3} include only the 
energy of nucleons, or should we add quark interactions?  Should we 
consider the kinetic energy of nucleons, or of the quarks and gluons
they comprise?  What if quarks themselves have constituents?

Clearly, a complete decomposition of the form \rref{a3} would require a
complete understanding of the physics of the materials we are studying.
Nevertheless, we can obtain important, if incomplete, information by 
isolating a few types of energy that are relatively well understood, making 
the assumption that there will be no precise cancellations between these 
contributions and those of the energies we ignore.  Without understanding 
the details of quantum chromodynamics, we cannot make sweeping statements 
about the equivalence principle for all binding energy; but we {\it can\/} 
draw conclusions about the particular contribution of the binding energy 
of nucleons in the atomic nucleus, which is understood empirically to a 
rather high accuracy.${}^{\citen{Leighton}}$  Similarly, when we consider 
kinetic energy below, we shall not attempt to analyze the kinetic energy of 
quarks, or even nucleons, but shall instead focus on the well-understood
physics of atomic electrons.  

Note that even for well-understood components of internal energy, the 
energy of a material is often not directly observable, but must be modeled 
theoretically.  We cannot, for example, dismantle the Earth to measure its 
gravitational binding energy, or catch a single atomic electron to measure 
its kinetic energy while it is still in the atom.  Often, however, there 
are useful internal checks on computed energies.  In particular, the virial 
theorem provides an important consistency check, and can sometimes be used 
to relate the energies in which we are interested to directly observable
quantities.  Recall that for a nonrelativistic bound system of $n$ 
particles at positions ${\bf r}_i$, having total kinetic energy $T$ and
interacting through a potential $U({\bf r})$, the virial theorem states 
that
\beq
\left\langle T \right\rangle 
  = {1\over2}\sum_{i=1}^n\left\langle {\bf r}_i\cdot\nabla U \right\rangle ,
\label{a4a}
\eeq
where the angle brackets denote a time average.  In particular, for an
electromagnetically bound system, $U({\bf r})\sim\sum q_i/|{\bf r} - 
{\bf r}_i|$, and it is not hard to show that
\beq
T = -U/2 = -E, 
\label{a4b}
\eeq
where $U$ is the electrostatic potential energy and $E$ is the total 
(kinetic plus potential) energy.  We shall use this relation below to 
check computed values of kinetic energy for atomic electrons.

Using an analysis based on equation \rref{a4}, physicists have obtained 
strong limits on violation of the equivalence principle by strong 
interaction energy in nuclei, electrostatic and magnetostatic energy in 
nuclei, the energy of hyperfine interactions, and weak interaction 
energy.${}^{\citen{Will}}$  The electrostatic binding energy of electrons 
in atoms has also been briefly discussed.${}^{\citen{Alvarez}}$  A 
somewhat different method, which uses Lunar laser ranging to compare 
the accelerations of the Earth and the Moon, has led to limits on 
violations of the equivalence principle by gravitational binding 
energy.${}^{\citen{Shapiro}}$

\section{The Case of Kinetic Energy}

Our interest in this article is kinetic energy, which has, surprisingly,
not yet been analyzed in the literature.  Most of the kinetic energy of 
an atom resides in the nucleus: a typical nucleus of atomic number 
$A$ has a radius $R\sim R_0A^{1/3}$ with $R_0\sim 1.3\times10^{-15}\,
\mathrm{m}$,${}^{\citen{Leighton}}$ and the uncertainty principle yields 
an estimate $T_{\mathit{nuc}}/m_Ic^2\sim 10^{-2}A^{-2/3}$.  This argument 
only sets a lower bound on the kinetic energy, however, and may not 
give the actual dependence on $A$; to use equation \rref{a4}, we would
need a much more sophisticated and model-dependent calculation.  For a 
system described by a simple enough potential, the virial theorem relates 
the kinetic energy to the (observable) binding energy.  In the nucleus, 
however, several types of potential energy compete, and the virial theorem 
does not give a unique separation of energies.  A decomposition like that 
of equation \rref{a3} is therefore problematic; we do not know enough to
distinguish the kinetic energy from other contributions.

For electrons in atoms, on the other hand, these problems largely 
disappear.  Accurate computations of electron kinetic energies are now
standard in condensed matter physics, and while exact solutions of the 
many-body Schr{\"o}dinger equation are not known, well-understood and 
well-tested approximations are readily available.  Moreover, atomic 
electrons are bound solely by electromagnetic interactions, and the 
virial theorem may be used to check computed kinetic energies against 
observed binding energies.  This simplicity comes at a price: electron 
kinetic energy is only a small part of total energy, 
$T_{\mathit{elec}}/m_Ic^2 \sim 10^{-7}$, 
and experiments are thus less sensitive to possible violations of the 
equivalence principle.  Nevertheless, the existing data are accurate 
enough to provide a good test in this relatively clean system.

The basic physics we wish to explore is fairly simple.  The electrons
in an atom with high atomic number are, on average, more tightly
bound than those in an atom with low atomic number.  Their kinetic
energy is consequently greater, and constitutes a greater proportion
of the total energy of the atom.  If, as an extreme example, this
kinetic energy had no weight, a high-$Z$ atom would fall measurably 
more slowly than a low-$Z$ atom.

To obtain quantitative predictions, we need to determine the kinetic
energy of atomic electrons.  We can begin with the simple Thomas-Fermi 
model for many-electron atoms,${}^{\citen{Landau}}$ which treats the 
electrons statistically as a Fermi gas and uses semiclassical methods
to determine their characteristics.  In this approximation, a typical 
electron in an atom of atomic number $Z$ is located roughly $Z^{-1/3}$ 
Bohr radii from the nucleus.  The electrostatic energy of one such 
electron is proportional to $Z/Z^{-1/3} = Z^{4/3}$; the energy of $Z$ 
electrons thus goes as $Z^{7/3}$.  By the virial theorem \rref{a4b}, 
the kinetic energy should have the same form, $T\sim cZ^{7/3}$.  
From the observed ground state energy of the hydrogen atom, we can 
estimate $c$ to be on the order of $10\,\mathrm{eV}$.  The ratios 
$T_{\mathit{elec}}/m_Ic^2$ in equation \rref{a4} thus range from about 
$10^{-6}$ for platinum to $10^{-8}$ for beryllium, much larger than the 
limits \rref{a2}--\rref{a2a} for violation of the equivalence principle.

To find more precise results, we could numerically integrate the 
Thomas-Fermi model to determine the coefficient $c$.  But it is almost 
as easy to employ the much more accurate numerical approaches that are
now widely used in condensed matter physics.  Consider, for example, the 
kinetic energies of atomic electrons in beryllium and copper.  These can 
be computed in the local density approximation${}^{\citen{Kohn}}$ (for a 
review of this method, see reference \citen{Jones}), using standard and 
widely available computer codes.  For isolated atoms, one finds that
\begin{eqnarray}
&&{T_{\mathit{elec}}\over m_Ic^2}({\mathrm{Be}}) = 4.6\times10^{-8} 
  \nonumber\\
&&{T_{\mathit{elec}}\over m_Ic^2}({\mathrm{Cu}}) = 7.7\times10^{-7} .
\label{a5}
\end{eqnarray}

To check these numbers, we can appeal to the virial theorem \rref{a4b} 
for electromagnetically bound systems, which allows us to compare the 
energies in equation \rref{a5} to published values of total energies as 
computed in the Hartree-Fock approximation${}^{\citen{Fraga}}$ and the 
local density approximation.${}^{\citen{Moruzzi}}$  The results agree to 
within 2\%.  Better yet, the total energy $E$ can be measured directly---it 
is the ionization potential, the energy required to totally ionize an 
atom---and the kinetic energies \rref{a5} can be compared to these 
observations.  We again obtain agreement to within 3\% for 
beryllium${}^{\citen{CRC}}$ and 1\% for copper.${}^{\citen{Sugar}}$  
One might worry that our computations were performed for isolated atoms, 
while the experimental test of reference \citen{EotWash} used solid 
metallic beryllium and copper.  But the relevant energy differences, the 
cohesive energies of the metals, are only a few eV per atom, completely 
negligible for our purposes.

We can now combine equations \rref{a2}, \rref{a4}, and \rref{a5}, assuming
as usual that no precise cancellations occur among different forms of 
energy.  We obtain a limit
\beq
|\eta_T|<6\times 10^{-6}
\label{a6}
\eeq
for violation of the equivalence principle by the kinetic energy of
electrons. 

A stronger, although less theoretically certain, limit can be obtained 
by comparing aluminum and platinum.  The local density approximation now 
gives
\begin{eqnarray}
&&{T_{\mathit{elec}}\over m_Ic^2}({\mathrm{Al}}) = 2.6\times10^{-7} 
  \nonumber\\
&&{T_{\mathit{elec}}\over m_Ic^2}({\mathrm{Pt}}) = 3.3\times10^{-6} .
\label{a7}
\end{eqnarray}
For aluminum, these figures are again in good agreement with the 
theoretical${}^{\citen{Fraga,Moruzzi}}$ and experimental${}^{\citen{Martin}}$
results for total energy.  For platinum, however, no experimental
results appear to be available, and the Hartree-Fock expression for 
total energy${}^{\citen{Fraga}}$ differs from the local density 
approximation for kinetic energy by about 15\%.  The difference may 
indicate a problem with the computed value; numerical errors in the 
code used for these calculations are likely to be significant for 
core electrons in high-$Z$ atoms.  In the absence of a better estimate 
of theoretical uncertainties, let us double this difference, and assume 
conservatively that the calculation \rref{a7} for platinum is accurate 
to within 30\%.  We can then combine equations \rref{a2a}, \rref{a4}, 
and \rref{a7} to obtain
\beq
|\eta_T|<6\times 10^{-7} .
\label{a8}
\eeq

The limits \rref{a6} and \rref{a8} are several orders of magnitude 
weaker than the corresponding results for nuclear energies.  Nevertheless, 
they are surprisingly strong, and may be counted as good evidence that 
the equivalence principle holds for kinetic energy.  

\section{General Covariance and the Weight of Light}

The results of the preceding section will come as no surprise to 
experts in relativity.  But perhaps they should.  We have another way 
of ``weighing'' kinetic energy: we can send a beam of particles past
a large mass (the Sun, say) and see how it is deflected.  It is well 
known that the deflection of light is twice that predicted by Newtonian 
theory; in this sense, at least, light falls with twice the acceleration
of ordinary ``slow'' matter.  

Indeed, the general relativistic deflection for a test particle with 
an arbitrary velocity $v$ and a large enough impact parameter $b$ 
is${}^{\citen{MTW,Koltun}}$
\beq
\theta = {2GM\over bv^2} \left( 1 + {v^2\over c^2}\right) .
\label{a8a}
\eeq
The corresponding angle in Newtonian gravity depends on the ratio of
gravitational to inertial mass, and it is easy to check that equation 
\rref{a8a} is just the Newtonian result for a particle with an inertial
mass $m_I$ and a gravitational mass
\beq
m_G = m_I \left(1 + {v^2\over c^2}\right)  
    = m_I + 2{T\over c^2} .
\label{a8b}
\eeq
For light, the kinetic energy $T$ in this expression should be replaced 
by the electromagnetic energy $U$, which can be loosely interpreted as
the ``kinetic energy'' of photons.

This argument does not seem to be widely published; instead, many 
texts rely on a simple ``Einstein elevator'' analysis that actually 
gives only half the correct deflection.${}^{\citen{Comer}}$  But it 
is not uncommon among students, and appears frequently in Internet 
discussions.  In such a simple form, the argument is easily addressed: 
it is attempting to impose Newtonian categories on general relativity, 
ignoring in particular the curvature of space.${}^{\citen{Ehlers}}$ 
To obtain a value for the deflection of light by the Sun we must at 
least implicitly compare measurements of direction in widely separated 
regions of space, and to perform such a comparison correctly, we need 
to take into account the curvature of space between these two 
regions.${}^{\citen{Will2}}$

There is a slightly more sophisticated version of this argument that is 
harder to dismiss, however.$^{\citen{Hughes}}$  Rather than sending a beam 
of light past the Sun, let us imagine confining the beam to a mirrored box 
near the Sun, thus avoiding the problem of comparing distant frames of
reference.  If the deflection \rref{a8a} truly reflects the ``weight of 
kinetic energy,'' a light beam with energy $U$ should contribute an amount 
$2U$ to the gravitational mass of the box.

We can analyze this situation in the weak field approximation to general 
relativity.${}^{\citen{Schutz}}$  In this limit, the metric is $g_{\mu\nu}
\approx \eta_{\mu\nu} + h_{\mu\nu}$, with
\beq
h_{00} = 2\phi, \qquad h_{ij} = 2\phi\delta_{ij} ,
\label{a9}
\eeq
where $\phi$ is the Newtonian gravitational potential.  The gravitational 
coupling to a test body with a stress-energy tensor $T^{\mu\nu}$ is 
thus\footnote{The coupling \rref{a10} can be obtained from the action
$S[\psi,g]$ for matter in curved spacetime by expanding the metric 
around its flat Minkowski value and noting that $\delta S = 
{1\over2}\sqrt{-g}\,T^{\mu\nu}\delta g_{\mu\nu}$.}
\beq
{1\over2}\int h_{\mu\nu}T^{\mu\nu} d^3x \approx
\int \phi\left(T^{00} + \delta_{ij}T^{ij} \right)d^3x .
\label{a10}
\eeq
For a slowly moving particle with rest mass $m$ and kinetic energy $T$,
we have
\beq
\int T^{00} d^3x \approx mc^2 + T, \qquad 
\int \delta_{ij}T^{ij}d^3x \approx 2T 
\label{a11}
\eeq
to lowest order in velocity.${}^{\citen{Weinberg}}$  For an electromagnetic 
field with energy $U$, on the other hand---or for a ``photon'' of energy 
$U$ moving at light speed---the stress-energy tensor is traceless, 
so${}^{\citen{Weinberg}}$
\beq
\int T^{00} d^3x \approx U, \qquad \int \delta_{ij}T^{ij}d^3x \approx U.
\label{a12}
\eeq
These expressions may be checked by considering a gas of particles (or 
photons) in a volume $V$, for which the diagonal spatial components $T^{ii}$ 
of the stress-energy tensor are equal to the pressure $p$.  In that case,
\rref{a11}--\rref{a12} may be recognized as the standard result that
\beq
pV = (\gamma-1)E
\label{a12a}
\eeq
where $E$ is the total (kinetic) energy and the coefficient $\gamma$ is
$5/3$ for nonrelativistic particles, $4/3$ for the extreme relativistic 
case.  

Our ``box of light'' consists of slowly moving walls with energies of the 
form \rref{a11} and a beam of light with energy of the form \rref{a12}.
For this system---or for any other system consisting of electromagnetic 
fields and matter---we thus have
\beq
\int T^{00} d^3x \approx mc^2+T+U, \qquad 
\int \delta_{ij}T^{ij}d^3x \approx 2T+U .
\label{a13}
\eeq
If $\phi$ is nearly constant over the system, equation \rref{a10} can 
thus be interpreted as a coupling of the Newtonian potential to the
combination $mc^2+3T+2U$.  For a free beam of light, the first two terms
are absent, and this analysis yields a gravitational deflection of twice 
the Newtonian value, as desired.  But comparing section \ref{sec1}, we seem 
to have found E{\"o}tv{\"o}s parameters $\eta_T=2$ and $\eta_U=1$, in gross 
violation of the equivalence principle.  

Of course, the coupling \rref{a13} does not really lead to a disagreement 
with experiment: we are saved by the virial theorem.  Equation \rref{a4b} 
was derived for a system of nonrelativistic charged particles coupled by
Coulomb interactions.  But the relativistic derivation of reference
\citen{Landau2} shows that the same relation holds for an arbitrary 
spatially bounded system of electromagnetically interacting particles
and electromagnetic radiation, as long as $U$ is now understood to be the 
total electromagnetic energy.  For our ``box of light,'' $2T+U$ therefore
vanishes, and $3T+2U=T+U=E$.  The apparent violation of the equivalence 
principle has thus rather mysteriously disappeared.  

Such an exact cancellation should have a fundamental explanation.  In 
general relativity, it is a consequence of general covariance.  Consider
a small electromagnetically bound ``test body'' in a gravitational field,
with center of mass ${\bar x}^i$.  Let us start with the expression 
\rref{a10} and perform a coordinate transformation
\beq
x^i \rightarrow {x^i}^\prime = x^i + \phi\xi(x) (x^i - {\bar x}^i)  ,
\label{a14}
\eeq
where $\xi$ is any function that is constant inside the test body and 
falls rapidly to zero outside.  In the weak field approximation, the fields 
$h_{\mu\nu}$ transform as${}^{\citen{Schutz}}$
\beq
h_{\mu\nu} \rightarrow h_{\mu\nu} 
  - \eta_{\mu\rho}\partial_\nu\delta x^\rho
  - \eta_{\nu\rho}\partial_\mu\delta x^\rho ,
\label{a14a}
\eeq
and it is easy to check that the coupling \rref{a10} becomes
\beq
\phi[mc^2 + T'+U' + (1-\xi(0))(2T'+U')] + \hbox{\it higher order terms} ,
\label{a15}
\eeq
where $T'$ and $U'$ are the kinetic and electromagnetic energies in the 
new coordinate system.  The coupling to $2T+U$ can thus be altered 
arbitrarily, and indeed ``gauged away,'' by a change of coordinates inside 
the test body.  This argument is closely related to our earlier appeal to 
the virial theorem: the relativistic virial theorem can be derived from 
conservation of the stress-energy tensor,${}^{\citen{Landau2}}$ which 
is in turn a consequence of general covariance in the weak field limit.

Put another way, we have learned that the determination of gravitational
mass from constituent energies is not entirely coordinate-independent.  
But the {\it physical\/} interaction with gravity {\it cannot\/} depend 
on a choice of coordinates: the coordinate-dependent part of the interaction 
must vanish.  That it does is guaranteed by the virial theorem, and indeed, 
this argument can be viewed as a derivation of the virial theorem.  We 
have also seen that within the framework of general relativity, coordinates 
can be chosen to show explicitly that the inertial and gravitational masses 
of a bound system are equal.    

We focused so far on electromagnetic energy.  But the coupling \rref{a15} 
illustrates a general ambiguity in determining the parameters $\eta_\alpha$.  
In any bound system, the virial theorem will require that some linear 
combination of energies vanish, and that some linear combination of the 
$\eta_\alpha$ therefore be unmeasurable.  For an electromagnetically 
bound system, for instance, a term in equation \rref{a3} of the form 
$\xi(2T+U)/c^2$ is inherently unobservable.

How can we deal with this ambiguity?  One answer is to compare systems 
with different interactions and different internal energies.  For any 
single system, the virial theorem will give a relationship among energies, 
and one combination $\sum c_\alpha\eta_\alpha$ will be undetermined.  But
the appropriate combination will differ from system to system.  Moreover, 
for systems with interactions described by nonpolynomial potentials, it
is evident from equation \rref{a4a} that the coefficients $c_\alpha$ will 
depend not only on the internal energies, but on the dynamics---the
average values of the positions ${\bf r}_i$---as well.  Observations 
of nuclei, for example, cannot detect a gravitational coupling to the 
combination $2T + U_{\mathit{elec}} + c U_{\mathit{nuclear}}$, where $c$ 
depends on mean distances between nucleons.  

If the coefficients $\eta_\alpha$ have any universal significance, we 
can now combine the limits coming from the energy content of nuclei with 
those coming from atomic electrons to obtain information about $\eta_T$ 
alone.  In particular, the electrostatic E{\"o}tv{\"o}s coefficient 
$\eta_U$ in nuclei satisfies${}^{\citen{Will}}$
\beq
|\eta_U|< 4\times10^{-10} ,
\label{a16}
\eeq
up to an ambiguous coupling that is {\em different\/} from the ambiguous
coupling for electrons.  It seems safe to assume that there should be 
no perverse cancellation between the gravitational couplings of, say,
electron kinetic energy in beryllium and nuclear binding energy in
platinum.   Equation \rref{a16} can thus be combined with our earlier
analysis to remove any ``virial theorem ambiguities'' in the limits 
\rref{a6} and \rref{a8} for kinetic energy.

We can thus tell our students with confidence that kinetic energy has
weight, not just as a theoretical expectation, but as an experimental
fact.  

\vspace{1.5ex}
\begin{flushleft}
\large\bf Acknowledgements
\end{flushleft}

The kinetic energies \rref{a5} and \rref{a7} were computed by Z.-W.\ Lu,
of the U.C.~Davis physics department; I am grateful for his help.  This 
work was supported in part by National Science Foundation grant PHY-93-57203 
and Department of Energy grant DE-FG03-91ER40674.

\end{document}